\documentclass[hyperref,showpacs,showkeys,12pt,titlepage]{revtex4}%
\usepackage{amsfonts}
\usepackage[debug,ps2pdf]{hyperref}
\usepackage{amsmath}
\usepackage{amssymb}
\usepackage[truetex]{graphicx}
\usepackage{bibmods}%
\providecommand{\U}[1]{\protect\rule{.1in}{.1in}}
\newtheorem{theorem}{Theorem}
\newtheorem{acknowledgement}[theorem]{Acknowledgement}

\begin{document}
\title{Influence of a Minimal Length on the Creation of Scalar Particles}
\author{S. Haouat}
\email{s.haouat@gmail.com}
\affiliation{\textit{LPTh, Department of Physics, University of Jijel, BP 98, Ouled Aissa,
Jijel 18000, Algeria.}}
\author{K. Nouicer}
\email{Khnouicer@yahoo.fr}
\affiliation{\textit{LPTh, Department of Physics, University of Jijel, BP 98, Ouled Aissa,
Jijel 18000, Algeria.}}

\begin{abstract}
In this paper we have studied the problem of scalar particles pair creation by
an electric field in the presence of a minimal length. Two sets of exact
solutions for the Klein Gordon equation are given in momentum space. Then the
canonical method based on Bogoliubov transformation connecting the
\textquotedblleft in\textquotedblright\ with the \textquotedblleft
out\textquotedblright\ states is applied to calculate the probability to
create a pair of particles and the mean number of created particles. The
number of created particles per unit of time per unit of length, which is
related directly to the experimental measurements, is calculated. It is shown
that, with an electric field less than the critical value, the minimal length
minimizes the particle creation. It is shown, also, that the limit of zero
minimal length reproduces the known results corresponding to the ordinary
quantum fields.

\end{abstract}
\maketitle

\section{Introduction}

The fact that strong electric field creates particle-antiparticle pairs from
the vacuum is predicted in the framework of quantum electrodynamics several
decades ago \cite{EH,Schwinger}. This effect, which is known as the Schwinger
effect, has a simple interpretation in the famous Dirac's hole theory - e.g.,
in the presence of an electric field, virtual particles can tunnel out of the
Dirac sea producing particle-hole pairs.

Since the publication of the seminal paper by Schwinger, a great interest is
devoted to the problem of particle creation from vacuum by strong fields.
Theoretically, the importance of this effect comes from its nonperturbative
nature and its relation with other problems such as the black hole radiation
and the dynamical Casimir effect. It is widely known, today, that the particle
creation effects have many important applications from heavy nucleus to black
hole physics \cite{Ruffini}.

In experimental physics, the strong field pair production has attracted much
attention, especially, in recent years. The field strength required to observe
produced pairs is of order of the critical value $E_{c}=\frac{m^{2}}%
{e}=10^{16}\mathtt{Vcm}^{-1}$ (for electron pairs), which seems to be beyond
the current technological capabilities. However, in recent years, explicit
experimental realizations have been proposed to see the Schwinger effect for
the first time \cite{exp1,exp2,exp3}. The basic principle of these experiments
is the enhancement of the Schwinger mechanism by the combination of a strong
slow pulsed laser with a weak fast pulsed laser. It is shown in \cite{exp1}
that the faster pulse gives a multi-photon contribution, which reduces the
barrier through which the particle tunnels and leads to an exponential
enhancement. Then the Schwinger effect could be observed in the near future.

Quantitatively, in $\left(  d+1\right)  $ dimensional space-time, the number
of created particles per unit of time per unit of volume is \cite{CF1}
\begin{equation}
\mathcal{N}=\frac{\left(  eE\right)  ^{\frac{d+1}{2}}}{\left(  2\pi\right)
^{d}}e^{-\pi\frac{m^{2}}{eE}}.
\end{equation}
The important characteristic of this formula is the exponential $e^{-\pi
\frac{m^{2}}{eE}}$, which explains the nonperturbative nature of the
phenomenon and the existence of the critical value from which the effect
becomes appreciable. Since this exponential is independent of the space-time
dimension, the analysis of the effect seems to be the same in arbitrary
dimensions. One expects, also, to obtain a similar exponential with a strong
slow pulsed laser because, in such a case, the period of the field is very
large compared to the typical time of the particle creation.

On the other hand, as is mentioned in \cite{Sabine}, there are many
indications that lead us to believe on the existence of a minimal length
scale. This minimal length scale, which is expected to be smaller than the
electroweak scale \cite{GUP1,GUP2,GUP3,GUP4,GUP5}, arises in many theories of
quantum gravity such as string theory \cite{ST1,ST2,ST3,ST4}, loop quantum
gravity \cite{LQG}, black hole physics \cite{BH1,BH2} and in non-commutative
field theories \cite{NC1,NC2,NC3}.

If such a minimal length exists in nature, it would be of great interest to
see how it influences the physical measurements. This explains why various
physical problems are reconsidered by taking into account the minimal length.
As example, we cite the harmonic oscillator \cite{Kempf4,HO1,H1}, the Hydrogen
atom \cite{H1,H2,H3,H4,H5,H6,H7}, the inverse square potential \cite{IS1}, the
Dirac oscillator \cite{Rel1}, and the resonant scattering by a potential
barrier \cite{RT,Salah}. Elsewhere, the influence of the minimal length on the
Casimir effect has been communicated in several works \cite{Casimir1,Casimir2}%
. We note, also, that the quantum corrections to the black hole thermodynamics
to all orders in the Planck length from a generalized uncertainty principle
are calculated in \cite{nouicer}. This kind of studies is motivated by the
possibility it offers to put the existence of a minimal length into evidence
and the regularization of certain problems in physics (see for instance
\cite{IS1,Ferkous}). Furthermore, as is mentioned in \cite{Nicolini}, since
the presence of a minimal length is common to many theories, phenomena such as
the Hawking effect and the particle creation, should be critically reviewed.

In this paper we propose to study the phenomenon of particle creation from
vacuum by an electric field in the (1+1) dimensional Minkowski space-time with
a nonzero minimal length. As in the case of the noncommutative space-time
\cite{Chair}, we expect that the introduction of a minimal length on the
theory of fields could have important consequences on the particle creation.
In addition, since the Schwinger effect is expected to be observed in the near
future, the minimal length could find an experimental justification through
this observation or at least find an important upper bound.

We consider in this paper the canonical method based on Bogoliubov
transformation connecting the "in" with the "out" states. In the first stage,
we give a short reminder about the particles creation problem and its
derivation from the wave functions both in position representation or in
momentum one. Then, we consider a scalar particle interacting with an electric
field in the presence of a minimal length, where we give two sets of exact
solutions for the corresponding Klein Gordon equation. In order to get the
good definition of the "in" and the "out" states, we study the limit of zero
minimal length. Next, we calculate the pair creation probability and the mean
number of created particles from the Bogoliubov coefficients. Finally, we
calculate the number of created particles per unit of time per unit of length
as soon as the imaginary part of the Schwinger effective Lagrangian.

\section{Usual theory of particle creation}

In order to derive the pair creation rate we have at our disposal several
methods such as the method based on vacuum to vacuum transition amplitude and
Schwinger-like effective action \cite{Schwinger,Itzykson}, the Hamiltonian
diagonalization technique \cite{Grib,Grib1}, the Feynman path integral method
\cite{Haw,Chit} as well as the semiclassical WKB approximation
\cite{Biswas1,Biswas2} and the "in" and "out" states formalism \cite{niki,PCC}
that we shall use in this work.

The "in" and "out" states formalism has been much used in the theory of
particle creation and vacuum instability in external fields. This formalism
proved most fruitful in finding the probability to create a pair of particles
and the mean number of created particles both in the presence of
electromagnetic fields or in curved space-time where gravitational fields are
present \cite{PCC}. However, the "in" and "out" states method is based on
analytic expressions of the wave functions which is not, in general, possible.
Since the constant electric field is described by a linear potential, the
corresponding Klein Gordon equation with minimal length admits exact and
analytic solutions only in momentum representation. Therefore, the
classification of these solutions as "in" and "out" states is not
straightforward. Thus, before considering the creation of scalar particles in
the presence of a minimal length, let us, first, recall briefly how to
determine the "in" and "out" states in ordinary quantum field theory by the
use of the momentum space. To our knowledge, there is no report, in
literature, on the definition of these states in momentum space. It is obvious
that the good definition of these states enables us to calculate the exact
probability of particle creation as soon as the mean number of created
particles with and without minimal length.

\subsection{Position representation}

In position representation the "in" and "out" states are well-defined. Here,
we briefly recall their definition. Starting from the (1+1) dimensional Klein
Gordon equation that describes the dynamics of a scalar matter field minimally
coupled to an external electric field%

\begin{equation}
\left[  \left(  \hat{p}_{\mu}-eA_{\mu}\left(  x\right)  \right)  ^{2}%
-m^{2}\right]  \psi\left(  t,x\right)  =0, \label{1}%
\end{equation}
where the 2-vector $A^{\mu}$ is given by%
\begin{equation}
\ A^{\mu}\equiv\left(  -Ex,0\right)  . \label{2}%
\end{equation}
We choose to work in natural units system where $\hbar=c=1$. Let us remark
that a constant electric field can be described by two straightforward gauges,
namely, the space-dependent gauge $A_{\mu}=(-Ex,0)$ and the time-dependent
gauge $A_{\mu}=(0,Et)$. In this work we consider the space-dependent gauge
because it seems simpler in the presence of a minimal length.

As is known, in order to solve the equation (\ref{1}), we write $\psi\left(
t,x\right)  =\exp(i\omega t)\varphi\left(  x\right)  $, where $\omega$ is the
energy of the particle. Then $\varphi\left(  x\right)  $ will be a solution of%
\begin{equation}
\left[  \left(  \omega+eEx\right)  ^{2}+\frac{\partial^{2}}{\partial x^{2}%
}-m^{2}\right]  \varphi\left(  x\right)  =0. \label{3}%
\end{equation}
By making the change
\begin{equation}
\xi=\sqrt{2ieE}\left(  x+\frac{\omega}{eE}\right)  \label{4}%
\end{equation}
we obtain the well-known differential equation%

\begin{equation}
\left[  \frac{\partial^{2}}{\partial\xi^{2}}-\frac{1}{4}\xi^{2}+\gamma
+\frac{1}{2}\right]  \tilde{\varphi}\left(  \xi\right)  =0, \label{5}%
\end{equation}
where $\tilde{\varphi}\left(  \xi\right)  \equiv\varphi\left(  x\right)  $ and%
\begin{equation}
\gamma=-\frac{1}{2}+i\frac{1}{2}\frac{m^{2}}{eE}. \label{6}%
\end{equation}
Equation (\ref{5}) admits two sets of exact solutions that can be written in
terms of Parabolic Cylinder Functions (PCFs) \cite{Grad}. According to
\cite{Hansen} and \cite{Greiner} the classification of these solutions as "in"
and "out" states is as follows%
\begin{align}
\varphi_{in}^{-}\left(  x\right)   &  =D_{\gamma^{\ast}}\left[  \left(
1-i\right)  \sqrt{eE}\left(  x+\frac{\omega}{eE}\right)  \right] \\
\varphi_{in}^{+}\left(  x\right)   &  =D_{\gamma}\left[  -\left(  1+i\right)
\sqrt{eE}\left(  x+\frac{\omega}{eE}\right)  \right] \\
\varphi_{out}^{-}\left(  x\right)   &  =D_{\gamma}\left[  \left(  1+i\right)
\sqrt{eE}\left(  x+\frac{\omega}{eE}\right)  \right] \\
\varphi_{out}^{+}\left(  x\right)   &  =D_{\gamma^{\ast}}\left[  -\left(
1-i\right)  \sqrt{eE}\left(  x+\frac{\omega}{eE}\right)  \right]  ,
\end{align}
Now, in order to determine the probability to create a pair of particles and
the mean number of created particles, we use the so called Bogoliubov
transformation connecting the "in" with the "out" states, which can be
obtained by taking into account that $\gamma^{\ast}=-\gamma-1$ and using the
formula \cite{Grad}%

\begin{equation}
D_{\gamma}\left(  \xi\right)  =\exp\left\{  i\pi\gamma\right\}  D_{\gamma
}\left(  -\xi\right)  -\frac{\sqrt{2\pi}}{\Gamma\left(  -\gamma\right)  }%
\exp\left\{  \frac{i\pi\gamma}{2}\right\}  D_{-\gamma-1}\left(  -i\xi\right)
. \label{11}%
\end{equation}
The relation between $\varphi_{in}^{\pm}$ and $\varphi_{out}^{\pm}$ reads%

\begin{align}
\varphi_{in}^{+}\left(  x\right)   &  =c_{1}\varphi_{out}^{+}\left(  x\right)
+c_{2}\varphi_{out}^{-}\left(  x\right) \label{12}\\
\varphi_{in}^{-}\left(  x\right)   &  =c_{2}^{\ast}\varphi_{out}^{+}\left(
x\right)  +c_{1}^{\ast}\varphi_{out}^{-}\left(  x\right)  , \label{13}%
\end{align}
where the Bogoliubov coefficients $c_{1}$ and $c_{2},$ given by
\begin{align}
c_{1}  &  =-\frac{\sqrt{2\pi}}{\Gamma\left(  -\gamma\right)  }\exp\left\{
\frac{i\pi\gamma}{2}\right\} \label{14}\\
c_{2}  &  =\exp\left\{  i\pi\gamma\right\}  , \label{15}%
\end{align}
fulfil the condition $\left\vert c_{1}\right\vert ^{2}-\left\vert
c_{2}\right\vert ^{2}=1$.

In quantum field theory, the relation between the "in" and the "out" modes
(\ref{12}) and (\ref{13}) can be converted into the following relation between
the creation and annihilation operators%

\begin{align}
a_{\omega,out}  &  =c_{1}\ a_{\omega,in}+c_{2}^{\ast}b_{\omega,in}%
^{+}\label{16}\\
b_{\omega,out}^{+}  &  =c_{2}\ a_{\omega,in}+c_{1}^{\ast}b_{\omega,in}^{+}.
\label{17}%
\end{align}
Therefore, the probability of pair creation and the mean number of created
particles will be given in terms of Bogoliubov coefficients. For instance, by
considering the probability amplitude%
\begin{equation}
\mathcal{A}=\left\langle 0_{out}\left\vert a_{\omega,out}b_{\omega
,out}\right\vert 0_{in}\right\rangle \label{18}%
\end{equation}
and by taking into account that\qquad%

\begin{equation}
b_{\omega,out}=\frac{1}{c_{1}^{\ast}}b_{\omega,in}+\frac{c_{2}^{\ast}}%
{c_{1}^{\ast}}a_{\omega,out}^{+} \label{19}%
\end{equation}
we obtain%

\begin{equation}
\mathcal{A}=\left\langle 0_{out}\left\vert a_{\omega,out}b_{\omega
,out}\right\vert 0_{in}\right\rangle =\frac{c_{2}^{\ast}}{c_{1}^{\ast}%
}\left\langle 0_{out}\mid0_{in}\right\rangle . \label{20}%
\end{equation}
As a result, the probability to create a pair of particles with the energy
$\omega$ from vacuum is given by%

\begin{equation}
\mathcal{P}_{\omega}=\left\vert \frac{c_{2}^{\ast}}{c_{1}^{\ast}}\right\vert
^{2}. \label{21}%
\end{equation}
Using the property%

\begin{equation}
\left\vert \Gamma\left(  \frac{1}{2}+iy\right)  \right\vert ^{2}=\frac{\pi
}{\cosh\left(  \pi y\right)  } \label{22}%
\end{equation}
we obtain the well-known result%
\begin{equation}
\mathcal{P}_{\omega}=\frac{\exp\left(  -\pi\frac{m^{2}}{eE}\right)  }%
{1+\exp\left(  -\pi\frac{m^{2}}{eE}\right)  }. \label{23}%
\end{equation}
This formulation enables us also to calculate the mean number of created
particles and the vacuum persistence. The mean number of created particles in
a state $\omega$ (the mean number of created particles per state) is defined
by the matrix element $n\left(  \omega\right)  =\left\langle 0_{in}\left\vert
a_{\omega,out}^{+}a_{\omega,out}\right\vert 0_{in}\right\rangle $, which can
be calculated to be%
\begin{equation}
n\left(  \omega\right)  =\left\vert c_{2}\right\vert ^{2}=\exp\left(
-\pi\frac{m^{2}}{eE}\right)  . \label{23b}%
\end{equation}
It should be noted that Eq. (\ref{23b}) is derived by considering the
commutators $\left[  \hat{a}_{\omega,out},\hat{a}_{\omega^{\prime},out}%
^{+}\right]  =\left[  \hat{b}_{\omega,out},\hat{b}_{\omega^{\prime},out}%
^{+}\right]  =\delta_{\omega\omega^{\prime}}$ with a finite time $T$ and
discrete values of the energy $\omega$. If we consider the limit
$T\rightarrow\infty$, the energy $\omega$ becomes continuous so that $n\left(
\omega\right)  $ can be interpreted as the number density of created particles
or the number of created particles per state.

Eqs. (\ref{23}) and (\ref{23b}) show that the present choice of "in" and "out"
states leads to the exact results of the particle creation. Let us show, in
the next paragraph, how these states can be expressed in momentum space.

\subsection{Momentum representation}

In momentum representation, the action of the operators $\hat{x}$ and $\hat
{p}$ is given by%
\begin{align}
\hat{p}  &  =p\label{24}\\
\hat{x}  &  =i\frac{\partial}{\partial p}, \label{25}%
\end{align}
and, therefore, the Klein Gordon equation can be written as
\begin{equation}
\left[  \left(  \omega+ieE\frac{\partial}{\partial p}\right)  ^{2}-p^{2}%
-m^{2}\right]  \tilde{\varphi}\left(  p\right)  =0 \label{26}%
\end{equation}
In order to solve this equation we factorize $\tilde{\varphi}\left(  p\right)
$ as follows
\begin{equation}
\tilde{\varphi}\left(  p\right)  =\exp\left(  i\frac{\omega}{eE}p\right)
F\left(  p\right)  . \label{27}%
\end{equation}
The new function $F\left(  p\right)  $ is then a solution of the following
equation
\begin{equation}
\left[  e^{2}E^{2}\frac{\partial^{2}}{\partial p^{2}}+p^{2}+m^{2}\right]
F\left(  p\right)  =0. \label{28}%
\end{equation}
By making the change%
\begin{equation}
\eta=\sqrt{\frac{2}{ieE}}p \label{29}%
\end{equation}
we obtain%
\begin{equation}
\left[  \frac{\partial^{2}}{\partial\eta^{2}}-\frac{1}{4}\eta^{2}+\gamma
^{\ast}+\frac{1}{2}\right]  f\left(  \eta\right)  =0, \label{30}%
\end{equation}
with $f\left(  \eta\right)  \equiv F\left(  p\right)  .$ We remark that the
solutions of equation (\ref{30}) can be written, also, in terms of PCFs.

Here, there is no reasonable criterion to find directly the good choice of
"in" and "out" states starting from the wave equation in momentum space. The
resort to the position space is then indispensable. Taking into account that
the passage from position space to momentum space can be realized by the
Fourier transformation, we obtain the following classification of the
solutions
\begin{align}
\tilde{\varphi}_{out}^{+}\left(  p\right)   &  =\exp\left(  i\frac{\omega}%
{eE}p\right)  D_{\gamma^{\ast}}\left[  \left(  1+i\right)  \sqrt{\frac{1}{eE}%
}p\right] \label{31}\\
\tilde{\varphi}_{out}^{-}\left(  p\right)   &  =\exp\left(  i\frac{\omega}%
{eE}p\right)  D_{\gamma}\left[  \left(  1-i\right)  \sqrt{\frac{1}{eE}%
}p\right] \label{32}\\
\tilde{\varphi}_{in}^{+}\left(  p\right)   &  =\exp\left(  i\frac{\omega}%
{eE}p\right)  D_{\gamma}\left[  -\left(  1-i\right)  \sqrt{\frac{1}{eE}%
}p\right] \label{33}\\
\tilde{\varphi}_{in}^{-}\left(  p\right)   &  =\exp\left(  i\frac{\omega}%
{eE}p\right)  D_{\gamma^{\ast}}\left[  -\left(  1+i\right)  \sqrt{\frac{1}%
{eE}}p\right]  . \label{34}%
\end{align}
For the calculation of the Fourier transformation of the PCFs one can use
equations (5.1) and (5.2) in \cite{Wolf}.

Then, by the use of (\ref{11}) we obtain the same Bogoliubov transformation as
in (\ref{12}) and (\ref{13})
\begin{align}
\tilde{\varphi}_{in}^{+}\left(  p\right)   &  =c_{1}\tilde{\varphi}_{out}%
^{+}\left(  p\right)  +c_{2}\tilde{\varphi}_{out}^{-}\left(  p\right)
\label{35}\\
\tilde{\varphi}_{in}^{-}\left(  p\right)   &  =c_{2}^{\ast}\tilde{\varphi
}_{out}^{+}\left(  p\right)  +c_{1}^{\ast}\tilde{\varphi}_{out}^{-}\left(
p\right)  . \label{36}%
\end{align}
Accordingly, this leads to the same probability and mean number as in
(\ref{23}) and (\ref{23b}). In reality, this result is unsurprising and it can
be predicted directly from (\ref{12}) and (\ref{13}). It shows, however, that
the Bogoliubov coefficients can be obtained directly from the "in" and the
"out" states expressed in momentum space. This will be very interesting in the
presence of the minimal length where there is no exact and analytic solution
in position space.

\section{Particles creation with minimal length}

Before considering the effect of the minimal length on creation of scalar
particles by an electric field, let us, first, give a short reminder about
this novel concept. There are many considerations that suggest the existence
of a minimal length. For example, in black hole physics \cite{Adler}, it is
shown that the Heisenberg uncertainty principle modifies to be
\begin{equation}
\Delta X\geq\frac{1}{2}\left(  \frac{1}{\Delta P}+\beta\Delta P\right)  ,
\label{eq:GUP}%
\end{equation}
where $\beta$ is very small positive parameter and $\hbar=1$. Such a
generalized uncertainty principle (GUP) leads to a nonzero minimal length
given by
\begin{equation}
\left(  \Delta X\right)  _{\min}=\sqrt{\beta}. \label{eq:MinLength}%
\end{equation}
In addition, since the relation (\ref{eq:GUP}) is invariant under the change
$\sqrt{\beta}\Delta P\rightarrow\frac{1}{\sqrt{\beta}\Delta P}$, there would
be a mixing between the ultraviolet and infrared behaviors of the field
theories, which is called the UV/IR mixing.

The derivation of this GUP by quantum mechanical tools in the one dimensional
space and its generalization to arbitrary dimensional case can be found in a
series of papers by Kempf and co-workers \cite{Kempf4,Kempf1,Kempf2,Kempf3}.
It is shown in these papers how the quantum mechanics with GUP can be
formulated. Here, we recall, briefly, that the GUP in (\ref{eq:GUP}) can be
reproduced by considering new operators $\hat{X}$ and $\hat{P}$ defined by
\begin{align}
\hat{X}  &  =\hat{x}\label{x}\\
\hat{P}  &  =f\left(  \hat{p}\right)  , \label{p}%
\end{align}
where $\hat{x}$ and $\hat{p}$ are the usual operators of quantum mechanics
fulfilling the commutation relation $\left[  \hat{x},\hat{p}\right]  =i$ and
$f\left(  \hat{p}\right)  $ is an injective function. The simpler case is to
consider the expansion%
\begin{equation}
f\left(  \hat{p}\right)  =\hat{p}\left(  1+\frac{\beta}{3}\hat{p}%
^{2}+...\right)  . \label{fp}%
\end{equation}
Taking into account that $\left[  \hat{X},\hat{P}\right]  =if^{\prime}\left(
\hat{p}\right)  $ and $\tilde{p}\approx\hat{P}\left(  1-\frac{\beta}{3}\hat
{P}^{2}+...\right)  $, we find that
\begin{equation}
\left[  \hat{X},\hat{P}\right]  =i\left(  1+\beta\hat{P}^{2}\right)  .
\label{CR1}%
\end{equation}
Then it is easy to show that this modified canonical commutation relation
leads to equation (\ref{eq:GUP}). It should be noted that the produced GUP
depends on the choice of the function $f$. The present choice is suitable for
a minimal length. In literature we find other choices that correspond to a
minimal length and/or maximal momentum (see for example \cite{pouria,pouria2}).

From equations (\ref{x}), (\ref{p}), (\ref{fp}) and (\ref{CR1}), we can derive
several representations for the operators $\hat{X}$ and $\hat{P}$. Among these
representations we quote the position representation defined by%
\begin{align}
\hat{X}  &  =x,\\
\hat{P}  &  =\left(  1+\frac{1}{3}\beta\hat{p}^{2}\right)  \hat{p},
\end{align}
with%
\begin{equation}
\ \hat{p}=-i\frac{\partial}{\partial x},
\end{equation}
and the momentum representation, where the action of the operators $\hat{X}$
and $\hat{P}$ is as follows%
\begin{align}
\hat{P}  &  =p,\label{eq:momentum}\\
\hat{X}  &  =i\left[  \left(  1+\beta p^{2}\right)  \frac{\partial}{\partial
p}\right]  . \label{eq:momentum2}%
\end{align}
Besides the fact that the present commutation relation provides us with exact
solvable model (see for example the works cited above), it offers the
possibility to show how the minimal length influences the physical
measurements. Then, to see the effect of this minimal length on the particle
creation, it is sufficient to consider equation (\ref{CR1}). The use of a more
general commutation relation risks leading to a nonsolvable problem and much
mathematical difficulties.

In the next paragraph we solve the Klein Gordon equation in the presence of an
electric field by taking into account the canonical commutation relation
(\ref{CR1}).

\subsection{Klein Gordon equation with a minimal length}

Before solving the Klein Gordon equation with a minimal length and studying
the creation of scalar particles, let us show that the minimal length theory
is consistent with the concept of gauge invariance. To this aim, we consider
the free Klein Gordon equation
\begin{equation}
\left(  -\frac{\partial^{2}}{\partial t^{2}}-\hat{P}^{2}-m^{2}\right)
\psi\left(  t,x\right)  =0, \label{free}%
\end{equation}
in the position space representation and we assume that a charged particle
described the Klein Gordon equation (\ref{free}) couples minimally to the
electromagnetic field following the general procedure%
\begin{align}
i\frac{\partial}{\partial t}  &  \rightarrow i\frac{\partial}{\partial
t}-eA_{0}\\
\text{\ }i\frac{\partial}{\partial x}  &  \rightarrow i\frac{\partial
}{\partial x}-eA_{1}.
\end{align}
Then, to the first order on $\beta$, the Klein Gordon equation becomes%
\begin{equation}
\left[  \left(  i\frac{\partial}{\partial t}-eA_{0}\right)  ^{2}-\left(
i\frac{\partial}{\partial x}-eA_{1}\right)  ^{2}-\frac{2\beta}{3}\left(
i\frac{\partial}{\partial x}-eA_{1}\right)  ^{4}-m^{2}\right]  \psi\left(
t,x\right)  =0. \label{gke}%
\end{equation}
Here, we remark that this equation is invariant under the gauge transformation%
\begin{align}
\psi^{\prime}\left(  t,x\right)   &  =e^{ie\alpha}\psi\left(  t,x\right)
\nonumber\\
A_{\mu}^{\prime}\left(  t,x\right)   &  =A_{\mu}\left(  t,x\right)
+\partial_{\mu}\alpha.
\end{align}
Let us, now, consider a scalar particle of mass $m$ and charge $e$ subjected
to a constant electric field $E$. In the presence of a minimal length the use
of the time-dependent gauge, with the assumption that $\psi\left(  t,x\right)
=e^{-ipx}\chi\left(  t\right)  $, leads to the following differential equation%
\begin{equation}
\left[  \frac{\partial^{2}}{\partial t^{2}}+\left(  p+eEt\right)  ^{2}%
+\frac{2\beta}{3}\left(  p+eEt\right)  ^{4}+m^{2}\right]  \chi\left(
t\right)  =0. \label{tgke}%
\end{equation}
To our knowledge there is no exact solution for this differential equation.
For this reason we consider the space-dependent gauge. With this choice
equation (\ref{gke}) reduces to
\begin{equation}
\left[  \frac{\partial^{2}}{\partial x^{2}}-\frac{2\beta}{3}\frac{\partial
^{4}}{\partial x^{4}}+\left(  \omega+eEx\right)  ^{2}-m^{2}\right]
\varphi\left(  x\right)  =0. \label{xgke}%
\end{equation}
$\allowbreak$The latter equation is of fourth order and, consequently, does
not admit exact and analytic solutions. Therefore, we use the momentum
representation, where the one dimensional stationary Klein Gordon equation reads%

\begin{equation}
\left[  \left(  \omega+eE\hat{X}\right)  ^{2}-\hat{P}^{2}-m^{2}\right]
\varphi=0, \label{40}%
\end{equation}
and the action of the operators $\hat{X}$ and $\hat{P}$ is shown in
(\ref{eq:momentum}) and (\ref{eq:momentum2}).

Then, in the $p$-representation, equation (\ref{40}) can be written as
\begin{equation}
\left[  e^{2}E^{2}\left(  1+\beta p^{2}\right)  ^{2}\frac{\partial^{2}%
}{\partial p^{2}}+2eE\left(  \beta eEp-i\omega\right)  \left(  1+\beta
p^{2}\right)  \frac{\partial}{\partial p}+p^{2}+m^{2}-\omega^{2}\right]
\tilde{\varphi}\left(  p\right)  =0. \label{41}%
\end{equation}
By making the change $p\rightarrow y,$ with
\begin{equation}
y=\frac{1-i\sqrt{\beta}p}{2} \label{42}%
\end{equation}
we obtain a Riemann type differential equation%
\begin{align}
&  \left[  \frac{\partial^{2}}{\partial y^{2}}+\left(  \frac{1-a_{1}%
-a_{1}^{\prime}}{y}-\frac{1-a_{3}-a_{3}^{\prime}}{1-y}\right)  \frac{\partial
}{\partial y}+\right. \nonumber\\
&  \left.  \left.  \left(  \frac{a_{1}a_{1}^{\prime}}{y}-a_{2}a_{2}^{\prime
}+\frac{a_{3}a_{3}^{\prime}}{\left(  1-y\right)  }\right)  \frac{1}{y\left(
1-y\right)  }\right]  g\left(  y\right)  =0\right.  \label{43}%
\end{align}
where the coefficients $a_{i}$ and $a_{i}^{\prime}$ are given by%

\begin{align}
a_{1}  &  =-\frac{\omega}{2eE\sqrt{\beta}}+i\frac{\sqrt{1-\beta m^{2}}%
}{2eE\beta}\label{44}\\
a_{1}^{\prime}  &  =-\frac{\omega}{2eE\sqrt{\beta}}-i\frac{\sqrt{1-\beta
m^{2}}}{2eE\beta}\label{45}\\
a_{3}  &  =\frac{\omega}{2eE\sqrt{\beta}}+i\frac{\sqrt{1-\beta m^{2}}%
}{2eE\beta}\label{46}\\
a_{3}^{\prime}  &  =\frac{\omega}{2eE\sqrt{\beta}}-i\frac{\sqrt{1-\beta m^{2}%
}}{2eE\beta}, \label{47}%
\end{align}
and%

\begin{equation}
a_{2}=1-a_{2}^{\prime}=\frac{1}{2}+\nu, \label{48}%
\end{equation}
with%
\begin{equation}
\nu=\frac{i}{eE\beta}\sqrt{1-\left(  \frac{eE\beta}{2}\right)  ^{2}}.
\label{49}%
\end{equation}
These coefficients satisfy the condition $a_{1}+a_{1}^{\prime}+a_{2}%
+a_{2}^{\prime}+a_{3}+a_{3}^{\prime}=1$.

Following \cite{Grad} equation (\ref{43}) admits two independent solutions
which can be written in terms of hypergeometric functions as follows
\begin{align}
\tilde{\varphi}_{_{1}}  &  =\left(  \frac{1-i\sqrt{\beta}p}{2}\right)
^{-\frac{\omega}{2eE\sqrt{\beta}}+i\frac{\sqrt{1-\beta m^{2}}}{2eE\beta}%
}\left(  \frac{1+i\sqrt{\beta}p}{2}\right)  ^{\frac{\omega}{2eE\sqrt{\beta}%
}+i\frac{\sqrt{1-\beta m^{2}}}{2eE\beta}}\nonumber\\
&  F\left(  \frac{1}{2}+\nu+i\frac{\sqrt{1-\beta m^{2}}}{eE\beta},\frac{1}%
{2}-\nu+i\frac{\sqrt{1-\beta m^{2}}}{eE\beta};1+i\frac{\sqrt{1-\beta m^{2}}%
}{eE\beta};\frac{1-i\sqrt{\beta}p}{2}\right)  \label{50}%
\end{align}
$\allowbreak$\linebreak and
\begin{align}
\tilde{\varphi}_{_{2}}  &  =\left(  \frac{1-i\sqrt{\beta}p}{2}\right)
^{-\frac{\omega}{2eE\sqrt{\beta}}-i\frac{\sqrt{1-\beta m^{2}}}{2eE\beta}%
}\left(  \frac{1+i\sqrt{\beta}p}{2}\right)  ^{\frac{\omega}{2eE\sqrt{\beta}%
}+i\frac{\sqrt{1-\beta m^{2}}}{2eE\beta}}\nonumber\\
&  F\left(  \frac{1}{2}+\nu,\frac{1}{2}-\nu;1-i\frac{\sqrt{1-\beta m^{2}}%
}{eE\beta};\frac{1-i\sqrt{\beta}p}{2}\right)  . \label{51}%
\end{align}
This choice of solutions, however, is not unique. If we use the fact that
equation (\ref{43}) is invariant under the change $y\rightarrow1-y$ and
$\left(  a,a^{\prime}\right)  \leftrightarrow\left(  c,c^{\prime}\right)  $,
we can find an other set of solutions $\left\{  \tilde{\varphi}_{_{3}}%
,\tilde{\varphi}_{4}\right\}  $, where
\begin{align}
\tilde{\varphi}_{_{3}}  &  =\left(  \frac{1+i\sqrt{\beta}p}{2}\right)
^{\frac{\omega}{2eE\sqrt{\beta}}+i\frac{\sqrt{1-\beta m^{2}}}{2eE\beta}%
}\left(  \frac{1-i\sqrt{\beta}p}{2}\right)  ^{-\frac{\omega}{2eE\sqrt{\beta}%
}+i\frac{\sqrt{1-\beta m^{2}}}{2eE\beta}}\nonumber\\
&  F\left(  \frac{1}{2}+\nu+i\frac{\sqrt{1-\beta m^{2}}}{eE\beta},\frac{1}%
{2}-\nu+i\frac{\sqrt{1-\beta m^{2}}}{eE\beta};1+i\frac{\sqrt{1-\beta m^{2}}%
}{eE\beta};\frac{1+i\sqrt{\beta}p}{2}\right)  \label{52}%
\end{align}
and%
\begin{align}
\tilde{\varphi}_{_{4}}  &  =\left(  \frac{1+i\sqrt{\beta}p}{2}\right)
^{\frac{\omega}{2eE\sqrt{\beta}}-i\frac{\sqrt{1-\beta m^{2}}}{2eE\beta}%
}\left(  \frac{1-i\sqrt{\beta}p}{2}\right)  ^{-\frac{\omega}{2eE\sqrt{\beta}%
}+i\frac{\sqrt{1-\beta m^{2}}}{2eE\beta}}\nonumber\\
&  F\left(  \frac{1}{2}+\nu,\frac{1}{2}-\nu;1-i\frac{\sqrt{1-\beta m^{2}}%
}{eE\beta};\frac{1+i\sqrt{\beta}p}{2}\right)  . \label{53}%
\end{align}
Thus, we have succeeded to find two sets of exact solutions for the Klein
Gordon equation with a minimal length in the presence of a constant electric
field. In the next paragraph we shall use these solutions to study the pair creation.

\subsection{The choice of "in" and "out" states and particle creation}

In order to classify our solutions as "in" and "out" states, we consider the
limit $\beta\rightarrow0$ and we compare the obtained solutions with the
results of section II (The calculation of this limit is shown in the
appendix). As a result, we find that the "in" states are given by
$\tilde{\varphi}_{1}\left(  p\right)  $ and $\tilde{\varphi}_{4}\left(
p\right)  $%
\begin{equation}%
\begin{array}
[c]{c}%
\tilde{\varphi}_{in}^{+}=\tilde{\varphi}_{1}\left(  p\right) \\
\tilde{\varphi}_{in}^{-}=\tilde{\varphi}_{4}\left(  p\right)
\end{array}
\label{54}%
\end{equation}
and the "out" states are given by $\tilde{\varphi}_{2}\left(  p\right)  $ and
$\tilde{\varphi}_{3}\left(  p\right)  $%
\begin{equation}%
\begin{array}
[c]{c}%
\tilde{\varphi}_{out}^{+}=\tilde{\varphi}_{2}\left(  p\right) \\
\tilde{\varphi}_{out}^{-}=\tilde{\varphi}_{3}\left(  p\right)  .
\end{array}
\end{equation}
Now, in order to obtain the relation between "in" and "out" modes and the
corresponding Bogoliubov coefficients, let us use the relation between
hypergeometric functions \cite{Grad}%
\begin{align}
F\left(  u,v;w;\xi\right)   &  =\frac{\Gamma\left(  w\right)  \Gamma\left(
w-v-u\right)  }{\Gamma\left(  w-u\right)  \Gamma\left(  w-v\right)  }F\left(
u,v;u+v-w+1;1-\xi\right)  ~~\nonumber\\
&  +\left(  1-\xi\right)  ^{w-u-v}\frac{\Gamma\left(  w\right)  \Gamma\left(
u+v-w\right)  }{\Gamma\left(  u\right)  \Gamma\left(  v\right)  }\nonumber\\
&  F\left(  w-u,w-v;w-v-u+1;1-\xi\right)  \label{55}%
\end{align}
to get%
\begin{equation}
\varphi_{in}^{+}=c_{1}\varphi_{out}^{+}+c_{2}\varphi_{out}^{-}%
\end{equation}
where, in this case the Bogoliubov coefficients are given by%
\begin{align}
c_{1}  &  =e^{i\vartheta}\frac{\Gamma\left(  \frac{1}{2}+\nu\right)
\Gamma\left(  \frac{1}{2}-\nu\right)  }{\Gamma\left(  \frac{1}{2}+\nu
+i\frac{\sqrt{1-\beta m^{2}}}{eE\beta}\right)  \Gamma\left(  \frac{1}{2}%
-\nu+i\frac{\sqrt{1-\beta m^{2}}}{eE\beta}\right)  }\label{57}\\
c_{2}  &  =\frac{\Gamma\left(  \frac{1}{2}+\nu\right)  \Gamma\left(  \frac
{1}{2}-\nu\right)  }{\Gamma\left(  1+i\frac{\sqrt{1-\beta m^{2}}}{eE\beta
}\right)  \Gamma\left(  -i\frac{\sqrt{1-\beta m^{2}}}{eE\beta}\right)  }
\label{58}%
\end{align}
where%
\begin{equation}
e^{i\vartheta}=-\frac{\Gamma\left(  i\frac{\sqrt{1-\beta m^{2}}}{eE\beta
}\right)  }{\Gamma\left(  -i\frac{\sqrt{1-\beta m^{2}}}{eE\beta}\right)  }.
\label{59}%
\end{equation}
By the use of (\ref{22}) and the following properties of Gamma functions
\begin{equation}
\Gamma\left(  x+1\right)  =x\Gamma\left(  x\right)  \label{61}%
\end{equation}
$\allowbreak$and%
\begin{equation}
\left\vert \Gamma\left(  iy\right)  \right\vert ^{2}=\frac{\pi}{y\sinh\left(
\pi y\right)  } \label{62}%
\end{equation}
we find%
\begin{equation}
\mathcal{P}_{\omega}=\frac{2\sinh^{2}\left(  \frac{\pi}{eE\beta}\sqrt{1-\beta
m^{2}}\right)  }{\cosh\left(  \frac{2\pi}{eE\beta}\sqrt{1-\beta m^{2}}\right)
+\cosh\left(  \frac{2\pi}{eE\beta}\sqrt{1-\left(  \frac{eE\beta}{2}\right)
^{2}}\right)  }. \label{60}%
\end{equation}
For the mean number of created particles, we have%
\begin{equation}
n\left(  \omega\right)  =\frac{\sinh^{2}\left(  \frac{\pi}{eE\beta}%
\sqrt{1-\beta m^{2}}\right)  }{\cosh^{2}\left(  \frac{\pi}{eE\beta}%
\sqrt{1-\left(  \frac{eE\beta}{2}\right)  ^{2}}\right)  }. \label{63}%
\end{equation}
Since the minimal length is supposed to be small we restrict our discussion to
the case when $\beta m^{2}\leq1$. It follows from equation (\ref{63}) that
when $\beta m^{2}=1$, the mean number $n\left(  \omega\right)  =0$. This is,
physically, plausible because particle-antiparticle pairs are defined only
when the measurable Compton wavelength of the particle is larger than the
minimal length.

\begin{figure}
[h]
\begin{center}
\includegraphics[
trim=0.000000in 0.000000in 0.000000in -0.178193in,
height=5.0215cm,
width=8.0309cm
]%
{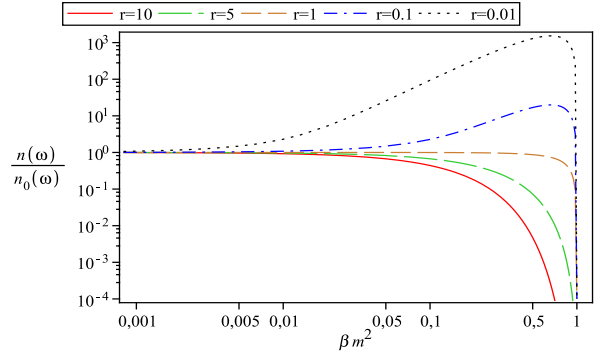}%
\caption{{\protect\small Plotting }$n\left(  \omega\right)  /n_{0}\left(
\omega\right)  $ {\protect\small as a function of the variable }$\beta m^{2}%
${\protect\small . The parameter }$r~${\protect\small is defined by }%
$r=\frac{E_{c}}{E}${\protect\small .}}%
\label{Fig 2}%
\end{center}
\end{figure}

In figure (\ref{Fig 2}), we plot the ratio $n\left(  \omega\right)
/n_{0}\left(  \omega\right)  $, where $n_{0}\left(  \omega\right)  $ is the
usual density of created particles (i.e. without minimal length) as a function
of the variable $\beta m^{2}$ for various values of $\frac{E_{c}}{E}\equiv r$.
As a result, we remark that the minimal length amplifies the scalar particle
creation when $\frac{E_{c}}{E}<1$ and minimizes it when $\frac{E_{c}}{E}>1$.
With the available technological capabilities the maximal strength of the
produced electric field is less than $E_{c}$ and consequently, the effect of
the minimal length is to reduce the scalar particle creation.%

%TCIMACRO{\FRAME{fhFU}{8.0309cm}{5.0215cm}{0pt}{\Qcb{\QTR{small}{Plotting
%}$n\left(  \omega\right)  /n_{0}\left(  \omega\right)  $ \QTR{small}{as a
%function of the variable }$\beta m^{2}$\QTR{small}{. The parameter
%}$r~$\QTR{small}{is defined by }$r=\frac{E_{c}}{E}$\QTR{small}{.}}}%
%{\Qlb{Fig 2}}{R1.eps}{\special{ language "Scientific Word";  type "GRAPHIC";
%maintain-aspect-ratio TRUE;  display "USEDEF";  valid_file "F";
%width 8.0309cm;  height 5.0215cm;  depth 0pt;  original-width 7.7089in;
%original-height 4.6164in;  cropleft "0";  croptop "1.0386";  cropright "1";
%cropbottom "0";  filename 'R1.eps';file-properties "XNPEU";}} }%
%BeginExpansion
%EndExpansion
As is shown in figure (\ref{Fig 2}), for small values $\beta m^{2}$, we see
that the minimal length can decrease or increase the pair creation rate. This
depends on the value of $\frac{E_{c}}{E}$. To put this effect into evidence,
let us take into account that $\beta$ is a small parameter and use the
approximations
\begin{equation}
\frac{2\pi}{eE\beta}\sqrt{1-\left(  \frac{eE\beta}{2}\right)  ^{2}}%
\approx\frac{2\pi}{eE\beta}-\pi\frac{eE\beta}{4} \label{65}%
\end{equation}
and%
\begin{equation}
\frac{2\pi}{eE\beta}\sqrt{1-\beta m^{2}}\approx\allowbreak\frac{2\pi}{eE\beta
}-\pi\frac{m^{2}}{eE}-\frac{2\pi}{eE}\frac{1}{8}m^{4}\beta. \label{66}%
\end{equation}
In such a case, the mean number of created particles by an electric field
takes the form
\begin{equation}
n\left(  \omega\right)  =\exp\left[  -\pi\frac{m^{2}}{eE}\left(  1+\frac{1}%
{4}\beta m^{2}\left(  1-\frac{e^{2}E^{2}}{m^{4}}\right)  \right)  \right]
\label{67a}%
\end{equation}
and the probability to create a pair of particles becomes%
\begin{equation}
\mathcal{P}_{\omega}=\frac{\exp\left[  -\pi\frac{m^{2}}{eE}\left(  1+\frac
{1}{4}\beta m^{2}\left(  1-\frac{e^{2}E^{2}}{m^{4}}\right)  \right)  \right]
}{1+\exp\left[  -\pi\frac{m^{2}}{eE}\left(  1+\frac{1}{4}\beta m^{2}\left(
1-\frac{e^{2}E^{2}}{m^{4}}\right)  \right)  \right]  }, \label{67b}%
\end{equation}
This means that a nonzero minimal length amplifies the scalar particle
creation when $\frac{E_{c}}{E}<1$ and minimizes it when $\frac{E_{c}}{E}>1$,
as is shown in figure (\ref{Fig 2}). In addition, we note that equations
(\ref{67a}) and (\ref{67b}) reduce, respectively, to (\ref{23}) and
(\ref{23b}), when $\beta\rightarrow0$.

Here, we have several remarks to add. As first remark, we note that we have
used, for the "in" and the "out" states in ordinary quantum field theory, the
choice of \cite{Hansen,Greiner}. The use of the choice of \cite{Niki1,hao}
leads to the same results.

Secondly, it should be noted that, since the Klein Gordon equation is gauge
invariant, one expects to obtain the same results by the use of the
time-dependent gauge. The proof of this can be done by considering the WKB
method starting from Eq. (\ref{tgke}). For instance, the pair creation
probability is given, in the WKB\ approximation, by
\begin{equation}
P\approx\exp\left[  -2\operatorname{Im}\oint\sqrt{m^{2}+\left(  p-eEt\right)
^{2}+\frac{2\beta}{3}\left(  p-eEt\right)  ^{4}}dt\right]  .
\end{equation}
Then by the use of the following integral%
\begin{equation}
\int_{0}^{b}\sqrt{\left(  x^{2}-b^{2}\right)  \left(  x^{2}-a^{2}\right)
}dx=\frac{a}{3}\left[  \left(  a^{2}+b^{2}\right)  E\left(  \frac{b}%
{a}\right)  -\left(  a^{2}-b^{2}\right)  K\left(  \frac{b}{a}\right)  \right]
,
\end{equation}
where $K\left(  k\right)  $ and $E\left(  k\right)  $ are, respectively, the
Euler integrals of first and second kind, and by taking into account that
$E\left(  k\right)  $ and $K\left(  k\right)  $ have the following expansions
for small values of $k$,
\begin{align}
K\left(  k\right)   &  =\frac{\pi}{2}\left(  1+\frac{1}{4}k^{2}+\frac{9}%
{64}k^{4}+...\right) \\
E\left(  k\right)   &  =\frac{\pi}{2}\left(  1-\frac{1}{4}k^{2}-\frac{3}%
{64}k^{4}+...\right)
\end{align}
we obtain%
\begin{equation}
P\approx\exp\left[  -\pi\frac{m^{2}}{eE}\left(  1+\frac{1}{4}\beta
m^{2}\right)  \right]  .
\end{equation}
This shows that the time dependent gauge gives approximately the same result
as the space dependent gauge.

Finally, let us remark that the quantity which is directly related to the
experimental measurements is the number of created particles per unit of time
per unit of length (in 1+1 dimensions). In the next subsection, we show how
this quantity can be calculated.

\subsection{The total number of created particles and the Schwinger effective
action}

Let us, first, calculate the total number of created particles by doing
summation over all states. The total number of created particles is given by%
\begin{equation}
N=\int\frac{T}{2\pi}d\omega~n\left(  \omega\right)  =\int\frac{dtd\omega}%
{2\pi}n\left(  \omega\right)  .
\end{equation}
Here, it should be noted that $\frac{dtd\omega}{2\pi}$ can be interpreted as
the number of states in the area sandwiched between the equal-energy contours
$\omega$ and $\omega+d\omega$, and the equal-time contours $t$ and $t+dt$ in
the $\left(  t,\omega\right)  $ plane. Therefore, the minimal length as
introduced in (\ref{CR1}) has no influence on the mesure $\frac{dtd\omega
}{2\pi}$ because the time and energy operators in the Klein Gordon equation
satisfy the ordinary canonical commutation relation.

According to \cite{Ni}, the integration over energy $\omega$ can be replaced
by%
\begin{equation}
\int d\omega=eE\int dx,
\end{equation}
where the variable $x$ denotes the position at which pairs with energy
$\omega$ are created. The total number of created particles is then given by%
\begin{equation}
N=\int dtdx\frac{eE}{2\pi}\frac{\sinh^{2}\left(  \pi\frac{\sqrt{1-\beta m^{2}%
}}{eE\beta}\right)  }{\cosh^{2}\left(  \frac{\pi}{eE\beta}\sqrt{1-\left(
\frac{eE\beta}{2}\right)  ^{2}}\right)  }. \label{1000}%
\end{equation}
On the other hand, if we write $N$ as%
\begin{equation}
N=\int dN=\int\frac{dN}{dtdx}dtdx, \label{1001}%
\end{equation}
we interpret $\frac{dN}{dtdx}\equiv\mathcal{N}$ as the number of created
particles per unit of time per unit of length. It follows from equations
(\ref{1000}) and (\ref{1001}) that
\begin{equation}
\mathcal{N}=\frac{dN}{dtdx}=\frac{eE}{2\pi}\frac{\sinh^{2}\left(  \pi
\frac{\sqrt{1-\beta m^{2}}}{eE\beta}\right)  }{\cosh^{2}\left(  \frac{\pi
}{eE\beta}\sqrt{1-\left(  \frac{eE\beta}{2}\right)  ^{2}}\right)  }.
\end{equation}
Here, we remark that for small values of $\beta$, $\mathcal{N}$ can be put in
the form
\begin{equation}
\mathcal{N}=\frac{eE}{2\pi}\exp\left[  -\pi\frac{m^{2}}{eE}\left(  1+\frac
{1}{4}\beta m^{2}\left(  1-\frac{e^{2}E^{2}}{m^{4}}\right)  \right)  \right]
. \label{appn}%
\end{equation}
This equation can be written as
\begin{equation}
\mathcal{N}=\mathcal{N}_{0}\left[  1-\frac{\pi}{4}\beta\frac{m^{4}}{eE}\left(
1-\frac{e^{2}E^{2}}{m^{4}}\right)  \right]  \label{N}%
\end{equation}
where $\mathcal{N}_{0}$ is the usual number of created particles per unit of
time per unit of length
\begin{equation}
\mathcal{N}_{0}=\frac{eE}{2\pi}\exp\left(  -\pi\frac{m^{2}}{eE}\right)
\end{equation}
and the factor $\left(  1-\frac{\pi}{4}\beta\frac{m^{4}}{eE}\left(
1-\frac{e^{2}E^{2}}{m^{4}}\right)  \right)  $ is the correction induced by the
minimal length.

If in future experiments the pair creation is observed, the relation (\ref{N})
could be used to quantify the predicted minimal length. Since the quantity
$\frac{m^{4}}{eE}$ is of order of $10^{-2}$ [GeV]$^{2}$ for $E=10^{-2}E_{c}$
and $m\approx m_{e}$, just the observation of the effect would imply that the
minimal length is smaller than $10^{-15}$m. Accurate measurements would lead
to an important upper bound.

Let us, now, show how the minimal length modifies the imaginary part of the
Schwinger effective action. It is well-known in quantum field theory that the
vacuum to vacuum transition amplitude can be expressed through an intermediate
effective action,
\begin{equation}
\left\langle 0_{out}\left\vert 0_{in}\right\rangle \right.  =\exp\left(
iS_{eff}\right)  =\exp\left(  i\int d^{2}x~\mathcal{L}_{eff}\right)  ,
\end{equation}
where $\mathcal{L}_{eff}$ is the Euler-Heisenberg effective Lagrangian
\cite{EH}. The probability of pair creation per unit of time and length can be
then extracted from the imaginary part of this Lagrangian%
\begin{equation}
\mathcal{P}_{Creat.}=\frac{1}{LT}\left[  1-\left\vert \left\langle
0_{out}\left\vert 0_{in}\right\rangle \right.  \right\vert ^{2}\right]
\simeq2\operatorname{Im}\mathcal{L}_{eff}.
\end{equation}
In the present case, it is easy to show that the vacuum persistence
$\mathcal{C}_{k}$ can be written in the form
\[
\mathcal{C}_{\omega}=1-\mathcal{P}_{\omega}=\frac{1}{1+\sigma},
\]
where%
\begin{equation}
\sigma=\frac{\sinh^{2}\left(  \pi\frac{\sqrt{1-\beta m^{2}}}{eE\beta}\right)
}{\cosh^{2}\left(  \frac{\pi}{eE\beta}\sqrt{1-\left(  \frac{eE\beta}%
{2}\right)  ^{2}}\right)  }.
\end{equation}
The vacuum to vacuum transition probability is then%

\begin{align}
\exp\left(  -2\operatorname{Im}S_{eff}\right)   &  =\prod_{\omega}%
\mathcal{C}_{\omega}\\
&  =\prod_{\omega}\exp\left[  -\ln\left(  1+\sigma\right)  \right] \\
&  =\exp\left[  -\sum_{\omega}\ln\left(  1+\sigma\right)  \right]
\end{align}
and, accordingly,%

\begin{equation}
2\operatorname{Im}S_{eff}=\int dxdt~2\operatorname{Im}\mathcal{L}_{eff}%
=\sum_{\omega}\ln\left(  1+\sigma\right)  .
\end{equation}
Here the symbol $\sum\limits_{\omega}$ denotes $\int\frac{dtd\omega}{2\pi}$.
Then the imaginary part of the effective Lagrangian can be written as
\begin{equation}
2\operatorname{Im}\mathcal{L}_{eff}=\frac{eE}{2\pi}\ln\left(  1+\frac
{\sinh^{2}\left(  \pi\frac{\sqrt{1-\beta m^{2}}}{eE\beta}\right)  }{\cosh
^{2}\left(  \frac{\pi}{eE\beta}\sqrt{1-\left(  \frac{eE\beta}{2}\right)  ^{2}%
}\right)  }\right)  .
\end{equation}
In the small $\beta$ limit, we get
\begin{equation}
2\operatorname{Im}\mathcal{L}_{eff}=\frac{eE}{2\pi}\ln\left\{  1+\exp\left[
-\pi\frac{m^{2}}{eE}\left(  1+\frac{1}{4}\beta m^{2}\left(  1-\frac{e^{2}%
E^{2}}{m^{4}}\right)  \right)  \right]  \right\}  . \label{64}%
\end{equation}
In addition, if we expand the logarithm function, we find the expression
\begin{equation}
2\operatorname{Im}\mathcal{L}_{eff}=\frac{eE}{2\pi}\sum_{n=1}\frac{\left(
-1\right)  ^{n+1}}{n}\exp\left[  -n\pi\frac{m^{2}}{eE}\left(  1+\frac{1}%
{4}\beta m^{2}\left(  1-\frac{e^{2}E^{2}}{m^{4}}\right)  \right)  \right]  ,
\end{equation}
which resembles to the well-know result corresponding to the scalar particle
creation in the (1+1) dimensional space-time \cite{CF1}, with the change
$m^{2}\rightarrow m^{2}\left(  1+\frac{1}{4}\beta m^{2}\left(  1-\frac
{e^{2}E^{2}}{m^{4}}\right)  \right)  $.

As is mentioned above, in the ordinary case, the exponential $\exp\left(  -\pi
m^{2}/eE\right)  $ appears in any $\left(  d+1\right)  $ dimensional
space-time. In the presence of the minimal length, we expect to have an
exponential similar to $\exp\left[  -\pi\frac{m^{2}}{eE}\left(  1+\frac{1}%
{4}\beta m^{2}\left(  1-\frac{e^{2}E^{2}}{m^{4}}\right)  \right)  \right]  $
in arbitrary dimensions.

\section{Conclusion}

In this paper, we have studied the problem of scalar particles pair creation
by an electric field in the presence of a minimal length by the use of the
canonical method based on Bogoliubov transformation. Although the
corresponding Klein Gordon equation is exactly solved in momentum space, it
was difficult to derive directly the pair creation probability. For this
reason, we have considered in the first stage the particle creation in
ordinary quantum field theory where we have written the "in" and the "out"
states in momentum representation. In the presence of a minimal length, we
have distinguished the "in" from the "out" states by studying the limit
$\beta\rightarrow0$. Then, we were able to extract the Bogoliubov coefficients
and to calculate the pair production probability and the mean number of
created particles. The number of created particles per unit of time per unit
of length, which is related directly to the experimental measurements, is calculated.

It is shown that the minimal length minimizes the particle creation when
$\frac{E_{c}}{E}\geq1$. This effect can be explained by the fact that, in the
presence of the minimal length, the threshold energy of the pair creation
modifies to be $2m\left(  1+\frac{1}{4}\beta m^{2}\left(  1-\frac{e^{2}E^{2}%
}{m^{4}}\right)  \right)  $ instead of $2m$. This could play a role in the
explanation why we do not see the assisted particle creation with the already
available technologies.

It is shown, also, that Schwinger mechanism can not create particles with mass
$m\sim\frac{1}{\sqrt{\beta}}$. Theoretically, this result could have a strong
impact on cosmology. If we reconcile that cosmological particle creation is
similar to the Schwinger effect, the creation of superheavy particles with the
mass of the Grand Unification scale in the early Universe, which is supposed
to have some important cosmological consequences \cite{SHP}, is then
suppressed by the GUP effects.

\appendix{}

\section{The $\beta\rightarrow0$ limit}

To find the good choice of "in" and "out" states let us study the limit
$\beta\rightarrow0$ that reproduces the ordinary case.

By using of the formula \cite{Grad}%

\begin{align}
F\left(  2a,2b;a+b+\frac{1}{2};\frac{1-z}{2}\right)   &  =A~F\left(
a,b;\frac{1}{2};z^{2}\right) \nonumber\\
&  +B~z~F\left(  a+\frac{1}{2},b+\frac{1}{2};\frac{3}{2};z^{2}\right)
\label{a1}%
\end{align}
where%
\begin{align}
A  &  =\frac{\Gamma\left(  a+b+\frac{1}{2}\right)  \Gamma\left(  \frac{1}%
{2}\right)  }{\Gamma\left(  a+\frac{1}{2}\right)  \Gamma\left(  b+\frac{1}%
{2}\right)  }\label{a2}\\
B  &  =\frac{\Gamma\left(  a+b+\frac{1}{2}\right)  \Gamma\left(  -\frac{1}%
{2}\right)  }{\Gamma\left(  a\right)  \Gamma\left(  b\right)  } \label{a3}%
\end{align}
and taking into account that%
\begin{align}
&  \left.  \lim_{\beta\rightarrow0}\left(  i\frac{\sqrt{1-\beta m^{2}}%
}{2eE\beta}-\frac{i}{2eE\beta}\sqrt{1-\left(  \frac{eE\beta}{2}\right)  ^{2}%
}\right)  =-i\frac{1}{4}\frac{m^{2}}{eE},\right. \label{a4}\\
&  \left.  \lim_{\beta\rightarrow0}\frac{\Gamma\left(  \frac{3}{4}%
+i\frac{m^{2}}{4eE}+\frac{i}{eE\beta}\right)  }{\Gamma\left(  \frac{1}%
{4}+i\frac{m^{2}}{4eE}+\frac{i}{eE\beta}\right)  }\sqrt{\beta}=\frac{1}%
{\sqrt{2eE}}\left(  1+i\right)  .\right.
\end{align}
and \cite{Grad}%
\begin{equation}
\lim_{v\rightarrow\infty}~~_{2}F_{1}\left(  u,v;w;\frac{z}{v}\right)
=~_{1}F_{1}\left(  u,w;z\right)  , \label{a5}%
\end{equation}
and by using the definition of PCFs \cite{Grad}
\begin{equation}
D_{\gamma}\left(  z\right)  =\frac{\sqrt{\pi}}{\Gamma\left(  \frac{1-\gamma
}{2}\right)  }F\left(  -\frac{\gamma}{2},\frac{1}{2};\frac{z^{2}}{2}\right)
-\frac{\sqrt{2\pi}}{\Gamma\left(  -\frac{\gamma}{2}\right)  }zF\left(
\frac{1-\gamma}{2},\frac{3}{2};\frac{z^{2}}{2}\right)  \label{a7}%
\end{equation}
we get%
\begin{equation}
\lim_{\beta\rightarrow0}\varphi_{_{1}}\left(  p\right)  \rightarrow
e^{i\frac{\omega}{eE}p}D_{\gamma}\left[  -\frac{\left(  1-i\right)  }%
{\sqrt{eE}}p\right]  . \label{a8}%
\end{equation}
With same steps we obtain for $\varphi_{3}\left(  p\right)  $%
\begin{equation}
\lim_{\beta\rightarrow0}\varphi_{3}\left(  p\right)  \rightarrow
e^{i\frac{\omega}{eE}p}D_{\gamma}\left[  \frac{\left(  1-i\right)  }{\sqrt
{eE}}p\right]  . \label{a9}%
\end{equation}
For $\varphi_{2}\left(  p\right)  $ and $\varphi_{4}\left(  p\right)  ,$ we
use in the first stage the property \cite{Grad}%
\begin{equation}
F\left(  a,b,c;y\right)  =\left(  1-y\right)  ^{c-a-b}F\left(
c-a,c-b;c;y\right)  , \label{a10}%
\end{equation}
and then we follow the same steps as for $\varphi_{1}\left(  p\right)  $ to
get%
\begin{equation}
\lim_{\beta\rightarrow0}\varphi_{2}\left(  p\right)  \rightarrow
e^{i\frac{\omega}{eE}p}D_{-\gamma-1}\left[  \frac{\left(  1+i\right)  }%
{\sqrt{eE}}p\right]  \label{a11}%
\end{equation}
and
\begin{equation}
\lim_{\beta\rightarrow0}\varphi_{4}\left(  p\right)  \rightarrow
e^{i\frac{\omega}{eE}p}D_{-\gamma-1}\left[  -\frac{\left(  1+i\right)  }%
{\sqrt{eE}}p\right]  . \label{a12}%
\end{equation}

\begin{acknowledgement}
We wish to thank the anonymous referees for their useful comments which
greatly improved the manuscript.
\end{acknowledgement}

\end{document}